# Absence of Anomalous Electron-Phonon Coupling in the Temperature Renormalization of the Gap of $CsPbBr_3$ Nanocrystals

Shima Fasahat[1], Benedikt Schäfer[1], Kai Xu[1], Nadesh Fiuza-Maneiro[2], Sergio Gómez-Graña[2], M. Isabel Alonso[1], Lakshminarayana Polavarapu[2], Alejandro R. Goñi[1,3*]

[1]Institut de Ciència de Materials de Barcelona, ICMAB-CSIC, Campus UAB, 08193 Bellaterra, Spain
[2]CINBIO, Universidade de Vigo, Materials Chemistry and Physics Group, Dept. of Physical Chemistry, Campus Universitario Lagoas Marcosende, 36310 Vigo, Spain
[3]ICREA, Passeig Lluís Companys 23, 08010 Barcelona, Spain

## Abstract

Metal halide perovskites exhibit a fairly linear increase of the bandgap with increasing temperature, when crystallized in a tetragonal or cubic phase. In general, both thermal expansion and electron-phonon interaction effects contribute equally to this variation of the gap with temperature. Herein, we have disentangled both contributions in the case of colloidal $CsPbBr_3$ nanocrystals (NCs) by means of photoluminescence (PL) measurements as a function of temperature (from 80 K to ambient) and hydrostatic pressure (from atmospheric to ca. 1 GPa). At around room temperature, $CsPbBr_3$ NCs also show a linear increase of the bandgap with temperature with a slope similar to that of the archetypal methylammonium lead iodide ($MAPbI_3$) perovskite. This is somehow unexpected in view of the recent observations in mixed-cation $Cs_xMA_{1-x}PbI_3$ single crystals with low Cs content, for which Cs incorporation caused a reduction by a factor of two in the temperature slope of the gap. This effect was ascribed to an *anomalous* electron-phonon interaction induced by the coupling with vibrational modes admixed with the Cs translational dynamics inside the cage voids. Thus, no trace of anomalous coupling is found in $CsPbBr_3$ NCs. In fact, we show that the linear temperature renormalization exhibited by the gap of $CsPbBr_3$ NCs is shared with most metal halide perovskites, due to a common bonding/antibonding and atomic orbital character of the electronic band-edge states. In this way, we provide a deeper understanding of the gap temperature dependence in the general case when the A-site cation dynamics is not involved in the electron-phonon interaction.

## INTRODUCTION

Colloidal lead halide perovskite nanocrystals (NCs) have attracted significant attention in recent years as promising candidates for next-generation optoelectronics [1-3] . This is due to their high photoluminescence (PL) quantum yields and the tunability of energy gap concerning different parameters including dimensionality, composition, as well as external stimuli like temperature and pressure [4-6]. Metal halide perovskites (MHPs) are materials with $ABX_3$ general formula, where A corresponds to an organic or inorganic cation like methylammonium (MA), formamidinium (FA) or Cs, B is a metallic bivalent cation such as Pb or Sn and X is a halide anion like I, Br, Cl. Mixed compositions in either A, B, or X positions provide optical properties tunability. As an example, mixed-anion compounds can be obtained via ionic exchange leading to the substitution of different halides [7]. This process is particularly effective in NCs [8]. Common structural behavior of these perovskites is that at high temperatures the crystal phase is cubic (C) and with decreasing temperature the phase transforms to tetragonal (T) and then orthorhombic (O) [9]. In the particular case of bulk $CsPbBr_3$, the phase transition from C to T takes place at around 400 K, and at around 360 K the transition from T to O phase occurs [10]. In the C and T phases, A-site cations are enclosed in $BX_6$ octahedral cages where they can freely move; depending on the chemical species this movement includes translation, rotation, and libration inside the cage voids. Recent studies highlighted the importance of the A-site cation dynamics at different temperatures and pressures in the structural behavior of MHPs [6, 9]. In particular, the fast roto-translational dynamics in cubic and tetragonal phases is fully or partially unfolded. In contrast, A-site cations are locked in specific positions and orientations inside the voids in less symmetric orthorhombic phases [11]. Regarding $CsPbBr_3$ NCs, the A-site cation dynamics also affects the structural behavior as compared to bulk. High-resolution transmission electron microscopy (TEM) studies of $CsPbBr_3$ single nanocubes at room temperature, in combination with results of simulations, hint at a supercooling of the cubic phase for NC sizes smaller than 10 nm [10]. That means that NCs are still cubic at temperatures much lower than the transition temperature for bulk (360 K), including room temperature. In contrast, in the case of larger NCs, they exhibit a minority of the O phase that coexists with the C [12].

A precise understanding of the band gap energy in semiconductors and its dependence on external parameters including temperature and pressure is essential for various optoelectronic applications. Regarding the behavior of the fundamental gap with temperature, most perovskite NCs in the tetragonal and cubic phase exhibit a linear dependence with a positive slope and a typical value of ca. 0.2 meV/K. Examples are $MAPbI_3$ [13-15], $MAPbBr_3$ [16-18], $FAPbI_3$ [14, 19, 20], $FAPbBr_3$ [13, 16, 21, 22], $CsPbI_3$ [13, 23-26], and $CsPbBr_3$ [13, 16, 23, 25, 27-29]. A notorious exception are $CsPbCl_3$ NCs and thin films [13, 25, 30], which display an outspoken negative slope around room temperature. In this respect, a quantitative analysis of the relative weight of thermal expansion and electron-phonon coupling on this temperature slope is still lacking, in particular for $CsPbBr_3$ NCs.

The energy gap dependence on temperature is bipartite. One point of consideration is the change of electronic band structure due to the variation of the lattice potential, which is the outcome of contraction/expansion of the lattice induced by temperature, known as the thermal expansion term (TE). The other contribution is the effect of lattice vibrations on the lattice potential that leads to energy renormalization of the electronic band structure, generally stronger at higher temperatures, called electron-phonon coupling term (EP). The variation of the gap concerning temperature reads as [31]:

$$\frac{dE_g}{dT} = \left[\frac{\partial E_g}{\partial T}\right]_{TE} + \left[\frac{\partial E_g}{\partial T}\right]_{EP} \qquad (1)$$

In the case of $MAPbI_3$, it was demonstrated that TE and EP terms possess almost equal weight in the temperature dependence of the gap [31]. In what follows, we will show, according to a literature survey that the linear temperature dependence of the gap in cubic and tetragonal phases is a general behavior for most of the perovskites. Nevertheless, there is a case in which a small amount of Cs incorporated in $MAPbI_3$ leads to a reduction of the slope by a factor of two [32]. This reduction was interpreted as due to an anomalous electron-phonon coupling, involving the interaction of electrons with the movement of the Cs inside the cage voids in synchrony with dynamic octahedral tilting. A motivation to investigate pure $CsPbBr_3$ NCs was to find out whether an anomalous behavior can be linked in general to Cs dynamics and to rationalize the different contributions to the temperature dependence of the gap in MPHs.

In this work, we studied the dependence of the energy gap of $CsPbBr_3$ NCs on temperature and pressure by means of PL experiments. We observe a linear temperature dependence of the energy gap of these NCs in the range around room temperature, for which the A-site cation dynamics (Cs in this case) is fully unfolded [11]. High-pressure experiments allowed us to disentangle TE and EP terms. This allows us to fully understand the temperature dependence of the bandgap. For the NCs, we observed a similar dependence as in bulk and other perovskites, with the TE and EP terms having similar magnitude as for $MAPbI_3$. Although we are dealing with pure Cs perovskites, there was no hint to an additional electron-phonon coupling mechanism that can be ascribed to the movement of the Cs and dynamic octahedral tilting whatsoever. In other words, we did not detect any anomalous electron-phonon coupling in the case of $CsPbBr_3$ NCs. On the contrary, $CsPbBr_3$ NCs show the common behavior of most MHPs, i.e. a linear gap temperature renormalization with a positive slope. This behavior, resulting from the bonding/antibonding character and the atomic orbital nature of the band-edge states, is considered normal in MHPs according to the survey carried out for the gap pressure and temperature coefficients and reported here.

## METHODS

### $CsPbBr_3$ NCs synthesis

In a typical synthesis, 15 mL of octadecene, 1.5 mL of oleic acid, and 1.5 mL of oleylamine and the precursor powders (1 mmol of $Cs_2CO_3$ and 3 mmol of $PbBr_2$) were loaded in a 50 mL glass vial. Then, the reaction medium was subjected to tip-ultrasonication (SONOPULS HD 3100, BANDELIN) for 30 min at a power of 30 W. During this time, a color change of the reaction mixture can be observed to yellow indicating the formation of $CsPbBr_3$ perovskite NCs. The solution was purified by centrifugation (10000 rpm, 10 min) and the pellet was redispersed in 6 mL of hexane. Finally, the solution was centrifuged again (5000 rpm, 10 min) to remove the large particles. The supernatant was collected to obtain $CsPbBr_3$ NCs. The produced NCs were analyzed by TEM using a FEI Tecnai G2 F20 HR(S) microscope.

### Temperature and Pressure-Dependent PL Experiments

For excitation of the PL spectra a violet laser (405 nm) was used, employing a very low laser power of ca. 2 μW (power density <15 W/cm$^2$) to prevent any photo-degradation of the samples [33, 34]. Spectra were recorded using a 20× long working distance objective with NA = 0.35 (laser spot of ca. 4 $\mu$m in diameter) and dispersed with a high-resolution LabRam HR800 grating spectrometer equipped with a charge-coupled device detector. PL spectra were corrected for the spectral response of the spectrometer by normalizing each spectrum using the detector and 600 grooves/mm grating characteristics. Temperature-dependent PL measurements on $CsPbBr_3$ NCs were carried out by decreasing the temperature between 300 and 80 K in steps of 5 K using liquid nitrogen in a gas flow cryostat from CryoVac with optical access that fits under the microscope of the Raman setup. The high-pressure PL measurements in the range up to ca. 1 GPa were performed at room temperature by employing a gasketed diamond anvil cell (DAC). Anhydrous propanol was used as a pressure-transmitting medium which ensures perfect hydrostatic conditions up to 4.2 GPa [35]. For loading the DAC a droplet of $CsPbBr_3$ NC solution was drop-casted on one of the diamonds of the DAC and allowed to dry. Then, a few ruby balls were added for pressure calibration [36]. After the pre-indentation, the thickness of the gasket was 120 $\mu$m. Afterward, a hole of ca. 250 $\mu$m was drilled with a sparkgap machine from EasyLab. This allowed us to adjust the pressure with the DAC in steps of less than 0.05 GPa, mainly at very low pressures (below 0.2 GPa). For this purpose, an electric motor drive was used to change the pressure in a continuous manner and at low speed (by ca. 0.05 GPa/min).

## RESULTS AND DISCUSSIONS

### Temperature-Dependent PL Measurements

The structural characterization of $CsPbBr_3$ NCs via high-resolution TEM indicates that the sample is fairly crystalline. Figures 1(a) shows a TEM image of $CsPbBr_3$ NCs with ordered cubic shapes and a relatively narrow size distribution with an average edge length of 8 nm, as depicted in Fig. 1(b).

The high resolution TEM images allowed us to measure the side lengths of the nanocubes precisely. The NC size was calculated by taking the average lateral length for each of the cuboids.

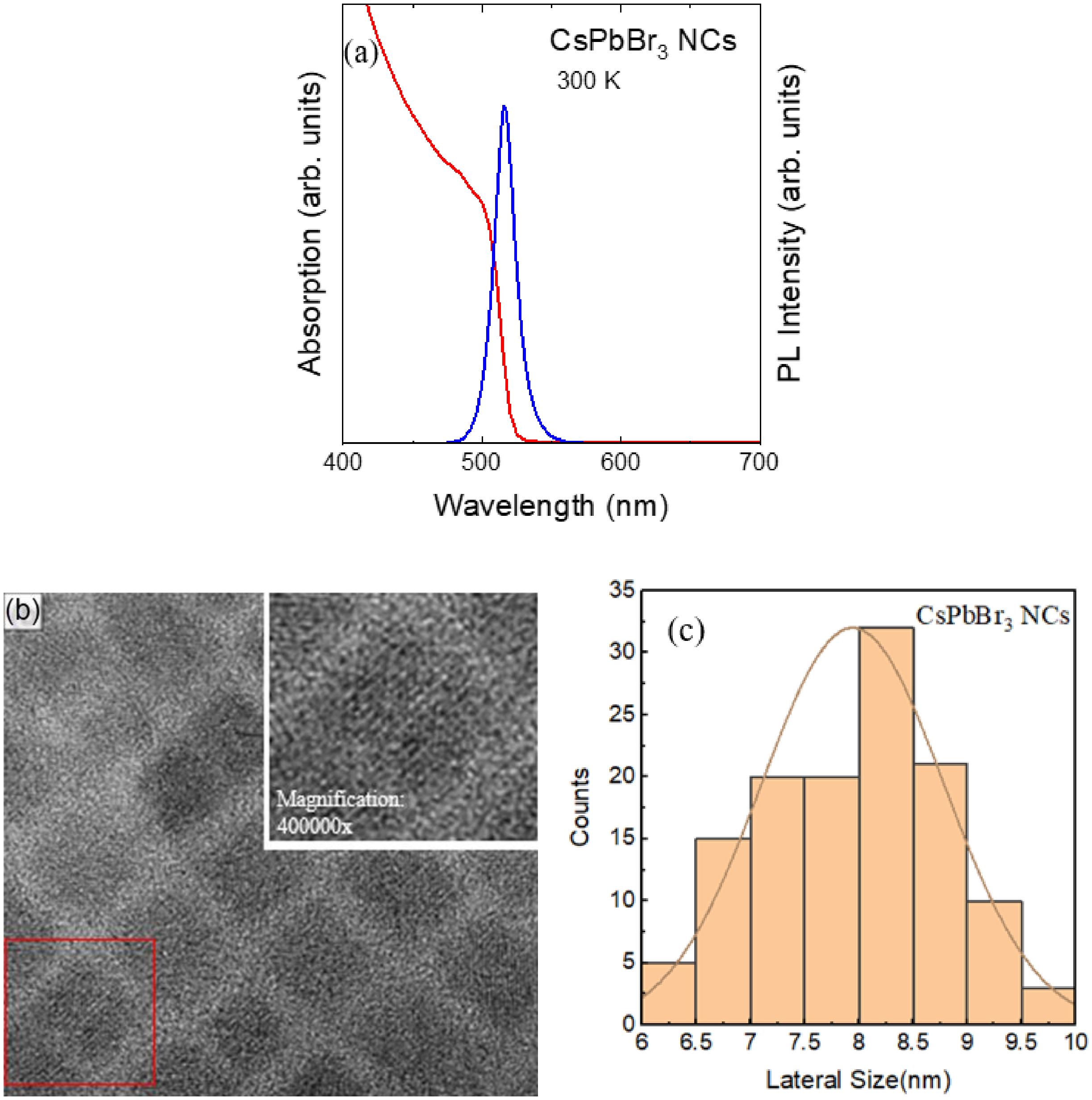

**Figure 1.** (a) Absorption and PL spectra of $CsPbBr_3$ NCs at room temperature. (b) High-resolution TEM micrograph of $CsPbBr_3$ NCs. The scale bar corresponds to 5 nm. (c) A representative histogram displaying the NCs size distribution was obtained by considering 126 $CsPbBr_3$ NCs.

As mentioned before, bulk $CsPbBr_3$ crystallizes at room temperature in an orthorhombic phase [10]. However, recent work on $CsPbBr_3$ nanocrystals (NCs) shows that the structural behavior of NCs is size-dependent. In fact, crystals exhibit a more cubic phase rather than orthorhombic when their size is smaller than 10 nm [12], similar to NCs used in this study. High-resolution TEM investigations of individual $CsPbBr_3$ nanocrystals have revealed that the transition temperature depends on their size. For nanocrystals smaller than approximately 10 nm, strain relaxation encourages the stabilization of the cubic α phase at room temperature when the Cs dynamics is unfolded [9, 10]. The Raman spectrum of $CsPbBr_3$ NCs (Fig. S1 of the Supporting Information) supports this statement. For completeness we point out that NCs produced by other synthetic methods such as ball milling [37] or hot injection [38] do crystallize in the orthorhombic phase. Nevertheless, the stabilization of the cubic phase down to temperatures much lower than the thermodynamic transition temperature has been reported in mixed-cation lead and tin iodide thin films [39]. Responsible for such a super-cooling of the cubic phase is the impact of thermally induced *dynamic* lattice distortions mediated by dynamic steric interaction, occurring at temperatures for which the A-site cation dynamics is fully unfolded [8]. In our case, we believe that the motion of the Cs cations inside the cage voids stabilize the cubic phase, at least, in the temperature range between 250 and 300 K (see discussion on the linearity range below).

Figure 2(a) displays the PL spectra of $CsPbBr_3$ NCs recorded at different temperatures from 80 to 300 K. These spectra were normalized to their absolute maximum intensity and vertically shifted to ease the comparison. Evolution of spectra in this temperature range exhibits an almost uniform behavior without phase transition [40]. The occasional small shift of the temperature-dependent spectra (within the width of the PL peak) is just due to slight inhomogeneities of the sample and the fact that the excitation of the PL does not always occur on the same spot. The main PL peak which is the result of the radiative recombination of free excitons [41], shows a small progressive shift to higher energies by increasing the temperature.

In order to analyze the PL spectra, a Gaussian-Lorentzian cross-product function was employed to describe the main PL peak. It is the same method utilized previously for analyzing PL peaks of $MAPbI_3$ [31, 42] , $Cs_xMA_{1-x}PbI_3$ (x=0.05, 0.1) perovskites [32], and a series of FA $_xMA_{1-x}PbI_3$ solid solutions [41, 43]. There are four adjustable parameters in this cross-product function which are: The amplitude prefactor *A*, the peak maximum position $E_0$, the full width at half maximum (FWHM) Γ, and a line shape weight *s* (0 for pure Gaussian, 1 for pure Lorentzian) [44]. PL peaks in these spectra have *s* values in the range of 0.4-0.6. Peak widths at temperatures between 80 K to 180 K are dominated by inhomogeneous (Gaussian) broadening, whereas from 185 K to 300 K the behavior of the line width reveals that homogeneous (Lorentzian) broadening takes gradually over. This is due to the size distribution and the small average size of the NCs. The spectra plotted in Fig. 2(a) show the red shift and sharpening of the peaks with decreasing temperature. The position of the maximum of the PL peaks $E_0$ can be considered to represent the energy of the bandgap. This is due to the fact that the exciton binding energy in this case is around 15 meV which is rather low [43]. The temperature dependence of the energy gap is plotted in Fig. 2(b). As mentioned previously, the behavior of $E_0$ with increasing temperature is quite steady with a total increase of almost 45 meV from 80 K to 300 K.

In this work, however, we will focus on the temperature range from approximately 250 to 300 K, in which $E_0$ exhibits a markedly *linear* temperature dependence with a positive slope, as indicated by the red line in Fig. 2b. The slope turns out to be $2.3(5)\times10^{-4}$ eV/K, a value similar to that previously obtained for $MAPbBr_3$ [18], $Cs_xMA_{1-x}PbI_3$ [32] and $FA_xMA_{1-x}PbI_3$ [41] in a similar temperature range around room temperature. The observation of such linearity in the temperature dependence of the gap is of fundamental and practical importance. On the one hand, the linear gap dependence is a consequence of the NCs being cubic and, viceversa, in the cubic phase, octahedral tilting plays no role and the temperature behavior of the perovskite gap is solely determined by the effects of bond stretching. The latter strictly depend on the bonding/antibond and atomic orbital character of the electronic states at the top of the valence band and the bottom of the conduction band, typically leading to a linear opening of the gap with increasing temperature [42,45]. Octahedral tilting, however, causes always an increase of the electronic gap, as a result of the effects of bond bending. The hybridization of the Pb 6*p* orbitals with the *s* orbitals of the corresponding halide anions imposes a serious constraint over the halide-Pb-halide bond to be straight. Any temperature (or pressure) induced tilting of the corner-sharing $PbX_6$ octahedrons gradually reduces the X-Pb-X bond angle below 180° with the concomitant blue shift of the gap energy [46-50]. Usually, the bond-bending effects (octahedral tilting) compete with those due to bond stretching, thus leading to an outspoken non-linear temperature or pressure dependence of the gap. Please notice that the *bowing* in the temperature dependence of the gap of the $CsPbBr_3$ NCs, setting in below ca. 250 K (see Fig. 2b), might be an indication of the structural phase transition from cubic to the thermodynamically favored orthorhombic phase [37, 38]. On the other hand, we note that in the whole linearity range, the first derivative of the gap over temperature is a constant, hence, both the TE and EP terms must be temperature independent as well.

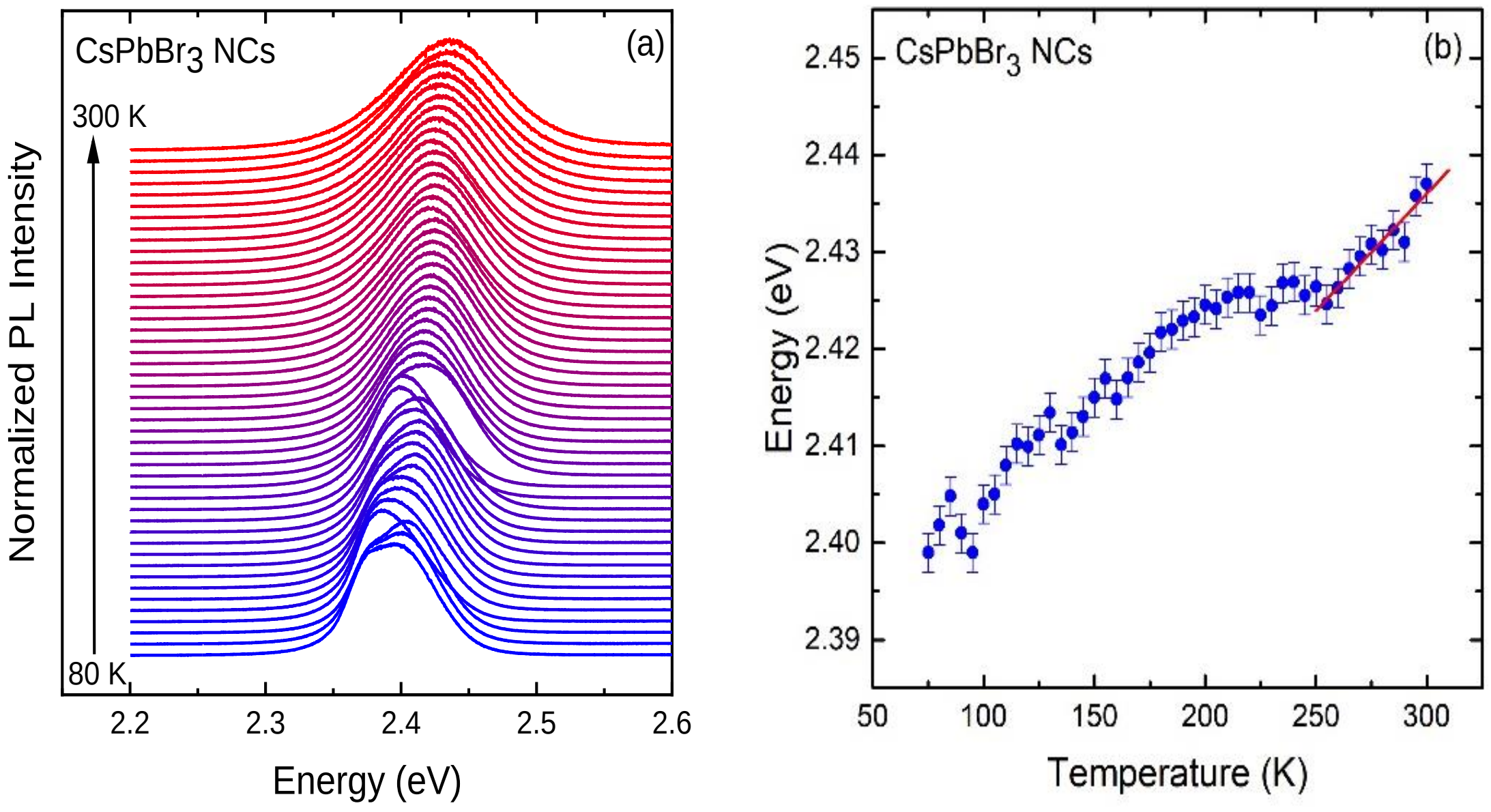

**Figure 2.** (a) PL spectra of $CsPbBr_3$ NCs recorded at various temperatures ranging from 80 to 300 K, in 5 K increments. The spectra were normalized to their peak intensity and vertically offset for better visualization. (b) Temperature dependence of energy gap of $CsPbBr_3$ NCs (blue symbols). The red line represents a linear fit to the data points around room temperature.

### Disentangling thermal expansion and electron-phonon coupling terms

As described in Eq. (1), two terms build the derivative of the energy gap concerning temperature: thermal expansion and electron-phonon coupling. The former can be identified with the effect of external hydrostatic pressure on the gap, which is directly related to that due to the shrinkage of the lattice with decreasing temperature. Thus, the following equation gives the thermal expansion term:

$$\left[\frac{\partial E_g}{\partial T}\right]_{TE} = -\alpha_v . B_0 . \frac{dE_g}{dP} \quad (2)$$

where $\alpha_v$ is the volumetric thermal expansion coefficient, $B_0$ refers to the bulk modulus and the last magnitude $\frac{dE_g}{dP}$ corresponds to the pressure coefficient of the energy gap, which will be assessed through high-pressure experiments [51]. Most direct band gaps in conventional semiconductors have a positive pressure coefficient at room temperature [45], therefore, according to Eq. (2), TE reduces the energy gap with increasing temperature. Lead halide perovskites, however, show the opposite behavior partly because the sign of the TE term is reversed in view of the fact that the sign of the gap pressure coefficient $\frac{dE_g}{dP}$ is negative [52].

PL spectra of $CsPbBr_3$ NCs were acquired at different pressures up to ca. 1 GPa (see Fig. S2 of the Supporting Information). As discussed previously, the center energy of the PL peaks is representative of the energy gap. Figure 3 shows the pressure dependence of the energy gap at room temperature. The straight red line drawn through data points is a linear fit and the slope gives directly the experimental value of the pressure coefficient, namely -0.055(15) eV/GPa. To calculate TE according to Eq. (2), we used values of volumetric expansion coefficient and bulk modulus reported in the literature as 1.14(10)$\times 10^{-4}$ $K^{-1}$ [53] and 21(3) GPa [54], respectively. Using these values, the obtained TE term amounts to 1.30(40)$\times 10^{-4}$ eV/K. This is almost half of the whole measured temperature slope of the gap. This means that the other half is related to the contribution that arises from the interactions between electrons and phonons, the EP term. Next, we analyze the contribution of EP coupling.

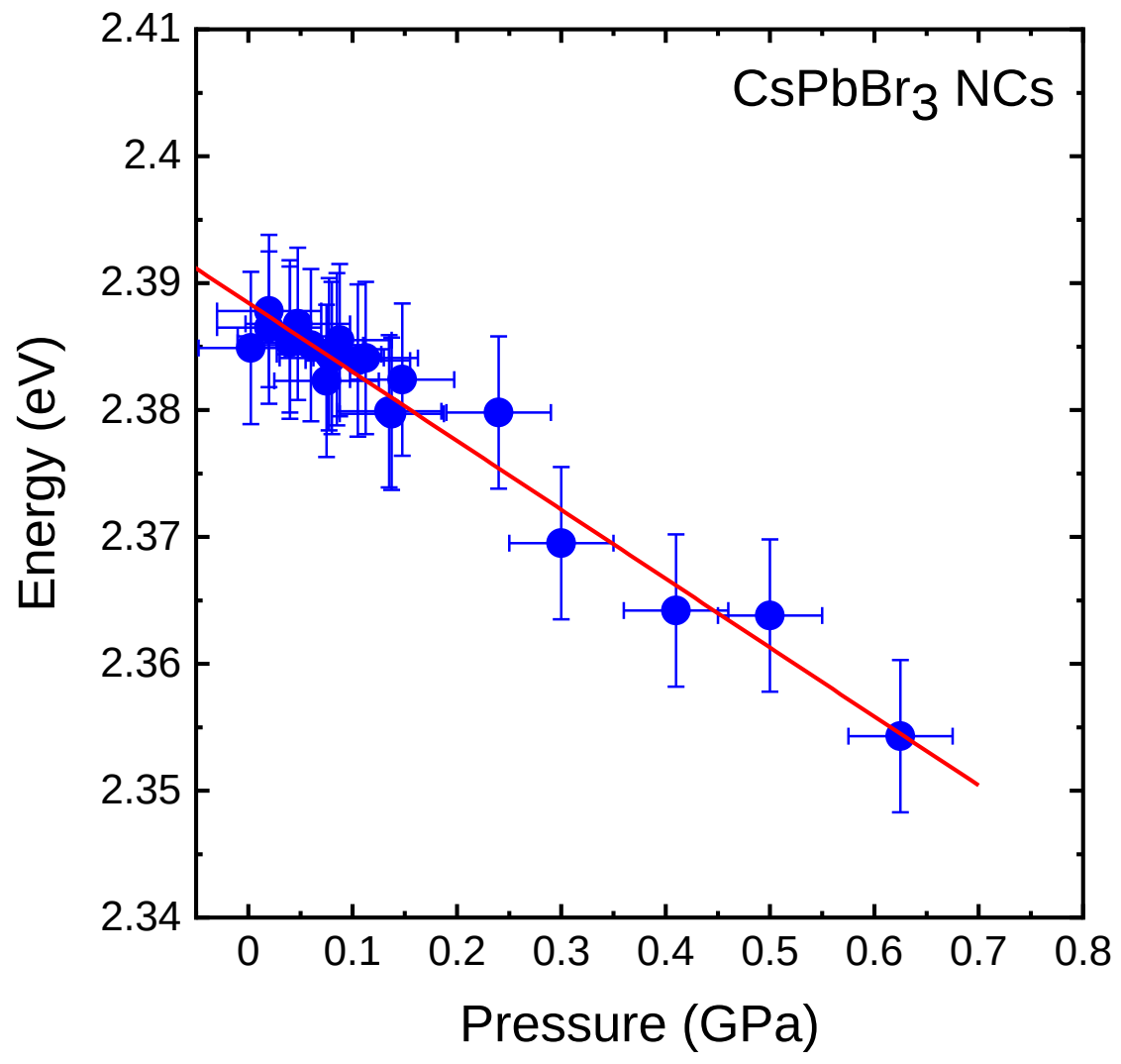

**Figure 3.** Pressure dependence of energy gap of $CsPbBr_3$ NCs recorded at room temperature.

The renormalization of the gap energy due to electron-phonon interaction is mainly infuenced by phonon modes corresponding to peaks in the phonon density of states (DOS) [55]. Indeed, this forms the basis of the Einstein oscillator model proposed by Cardona and co-workers. [56-58], to describe the EP term in a practical way by considering effective electron-phonon interaction coefficients $A_i$ for phonons with an average frequency $\omega_i$, derived from peaks in the phonon DOS. The electron-phonon correction to the gap reads consequently as:

$$\left[\Delta E_g(T)\right]_{EP} = \sum_i A_i \cdot \left(n_B(\omega_{i,}T) + \frac{1}{2}\right) \quad (3)$$

Here, $n_B$ denotes the Bose-Einstein phonon occupation factor and shows the explicit temperature dependence of the renormalization. The coefficient $A_i$ is temperature-independent but it may be different depending on the particular average frequency and the mass of the atomic species involved in the vibration. This specific aspect can be taken into account in materials with two atoms per unit cell that have significantly different masses, such as the cuprous halides, by using a two-oscillator model [57, 58]. In that case, this is justified by the phonon DOS showing two peaks: one at the average frequency of the acoustic phonon branches at the Brillouin zone edges (corresponding to vibrations of the heavier atom), and another representing the optical phonon contribution (corresponding to vibrations of the lighter atom). However, this scheme cannot be directly applied to halide perovskites, which have multiple atoms per unit cell and peaks in the DOS due to mixed character vibrations that cannot be simply categorized by lead or halide atoms alone. Despite this more complicated structure, the simple Einstein oscillator concept using just one average frequency was shown to be a good approximation in MHPs [31]. Hence, the electron-phonon renormalization can be effectively modeled using a single Einstein oscillator with a positive amplitude, $A_{eff}$, to replicate the linear reduction of the gap with decreasing temperature in $CsPbBr_3$ NCs. Due to the unknown absolute magnitude of the gap renormalization at room temperature (or any other temperature), we consider instead the temperature derivative of the gap as follows:

$$\left[\frac{\partial E_g}{\partial T}\right]_{EP} = \frac{A_{eff}}{4T} \cdot \frac{\hbar\omega_{eff}}{k_B T} \cdot \frac{1}{sinh^2\left(\frac{\hbar\omega_{eff}}{2k_B T}\right)} \quad (4)$$

In this way, all the crucial elements to evaluate the weight of electron-phonon coupling in the band gap renormalization are available. Figure 4a presents smoothed data related to PL peak energies as a function of temperature. Data smoothing was performed to prevent undesired amplification of the scatter of the data points during the numerical calculation of the first derivative of $E_0$ with respect to temperature. The latter (numerical derivative of experimental $E_0$ values) are displayed in Fig. 4(b) as green circles. Only the filled green circles corresponding to the stability range of the cubic phase are considered during the fitting process (linearity regime). The blue dash-dotted curve in Fig. 4(b) represents the contribution of TE already found as a constant value of $1.30(40)\times 10^{-4}$ eV/K. Taking this value into account together with the function in Eq. (4), a fit to the experimental points was performed in the selected temperature range of 250 K to 300 K. The adjustable parameters in the EP term determined for $CsPbBr_3$ NCs are $A_{eff}$ = 6.7(5) meV for the electron-phonon coupling amplitude and $\hbar\omega_{eff}$ = 6(1) meV for the oscillator frequency. The result of the least-squares fit is represented by the black curve. Despite the slight dispersion in the first derivative values, the average slope obtained from the fit (black line) is $2.24(2)\times10^{-4}$ eV/K, in excellent agreement with the slope of the red curve in Fig. 4a.

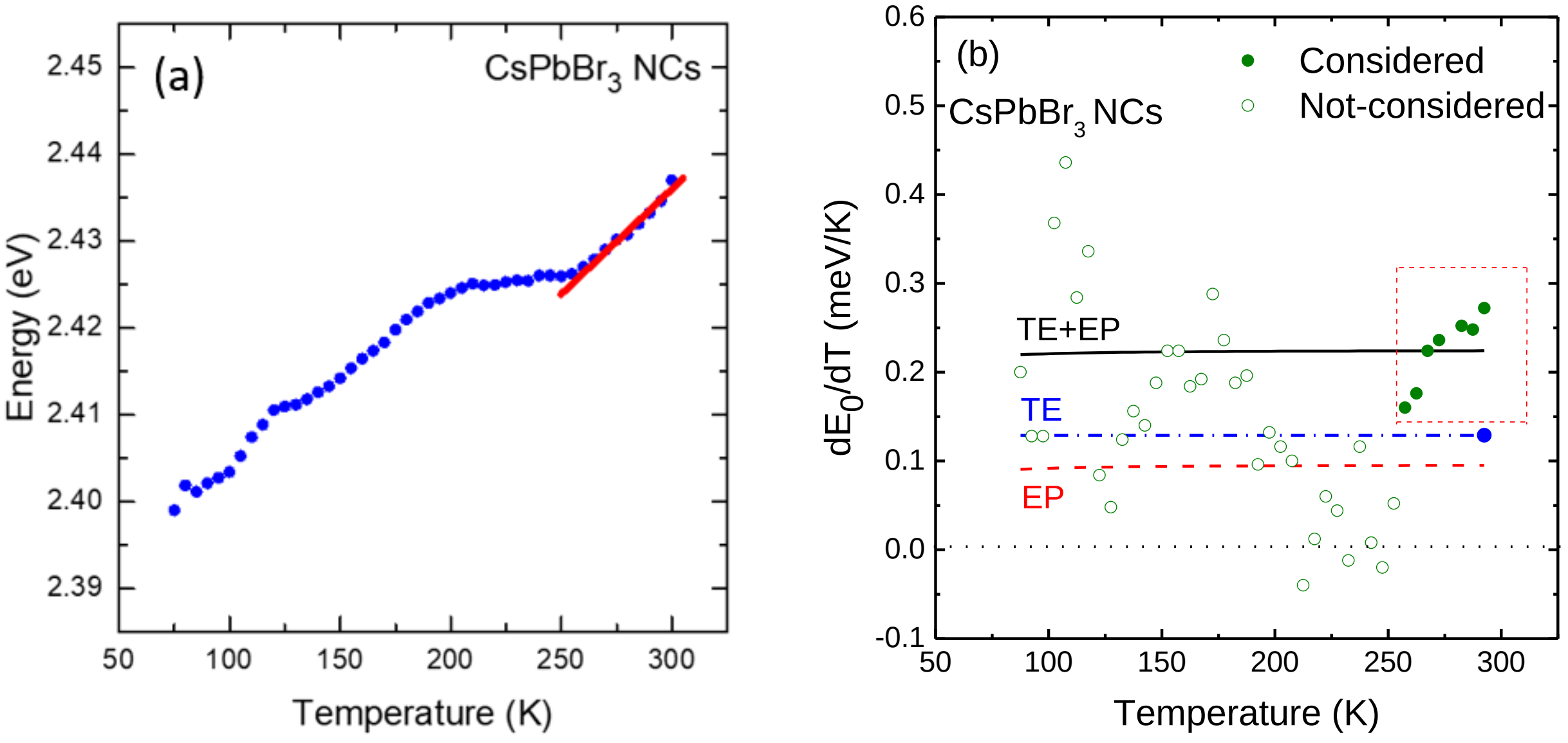

**Figure 4.** (a) Smoothed data related to the maximum PL peak energies for $CsPbBr_3$ NCs plotted as a function of temperature (blue circles) and linear dependence using the fitted slope (red line) around room temperature. (b) First derivative of the PL peak energy $E_0$ with respect to temperature (green solid and open symbols), numerically obtained from the smoothed data of (a). Only the points shown in the red rectangle (linearity regime) are considered. The different contributions to the band gap renormalization are indicated.

While the contribution of the EP term near room temperature is clear from the experimental data, its interpretation in terms of Eq. (4) required to discriminate between multivalued solutions due to the fact that several $A_{eff}$ and $\hbar\omega_{eff}$ value pairs can yield the same fit quality. We chose to fix the value of the effective frequency of the Einstein oscillator to 6(1) meV whereas the amplitude $A_{eff}$ was kept free in the fitting process. The fixed value of the effective frequency is actually the one previously obtained for $MAPbI_3$ [28]. In fact, if we look at the phonon DOS for $MAPbI_3$, the electron-phonon coupling term accounts only for the coupling with the phonons of the inorganic cage [34]. In addition, we note that this value is also close to the effective phonon frequency involved in the exciton broadening as a function of temperature in eleven mixed cation $FA_xMA_{1-x}PbI_3$ solid solutions [27]. All this supports the idea that the frequency value selected here is quite representative of the EP coupling to the inorganic cage phonons, hence holding for most of the perovskites.

Despite dealing with pure $CsPbBr_3$ NCs, the fact that the EP term is fully accounted for with a single Einstein oscillator representing the inorganic cage phonons means that there is no hint to an anomalous EP coupling due to Cs dynamics. This is in frank contrast to the mentioned case of $Cs_xMA_{1-x}PbI_3$ single crystals (x = 0.05 and 0.1), where Cs substitution leads to an anomalous EP due to the indirect coupling between the charge carriers of the inorganic cage and the fast translational dynamics of the Cs cations [32]. This anomalous indirect coupling occurs when the dynamic tilting of the PbI6 octahedra occurs in synchrony with the motion of the Cs cations between the potential minima within the cage voids. In the case of the $CsPbBr_3$ NCs, dynamic octahedral tilting fluctuations centered at zero tilt, as is mandatory for a cubic phase. Consequently, although the dynamics of the Cs cation is unfolded, the motion is restricted to the central region of the cage voids and is characterized by a spherical atomic probability density cloud. In this situation, coherence is simply forbidden by symmetry, so only normal EP coupling remains but no anomalous EP coupling.

We now will show that the behavior observed here, featuring a *normal* EP coupling term, is general and holds for most of the MPHs. The pressure and temperature dependence of the gap of different MPHs is summarized in Table 1. We list experimental results on the temperature and pressure coefficient of the fundamental gap obtained for lead halide perovskites with formula $APbX_3$ in single, poly and nanocrystalline form as well as thin films. It is important to note that the data are circumscribed to near ambient conditions, for which most of the materials crystallize either in a cubic or tetragonal phase. This corresponds here to the linearity regime considered for the analysis of the TE and EP contributions to the gap renormalization. In these phases, static octahedral tilting does not play any role and the gap varies linearly with temperature. Otherwise, strong non-linearities would affect the temperature and pressure behavior of the perovskite gap. Close inspection of the tabulated data allows for drawing a few interesting conclusions. On one hand, all pressure coefficients are negative, without exceptions, with values bunching between ca. -30 and -70 meV/GPa. On the other hand, except for $CsPbCl_3$, the linear temperature coefficients are positive and in the range of 0.15 and 0.4 meV/K.

In the following, it will be shown that both trends are not fortuitous but a direct consequence of the bonding/antibonding and atomic orbital character of the electronic states at the top and bottom of the valence and conduction band, shared by all the tabulated materials. Regarding the gap pressure coefficient, it has been shown for the archetypal perovskite $MAPbI_3$ [42] that the pressure-induced red shift of the gap in the tetragonal phase, stable at ambient conditions, can be accounted for in terms of the well-established systematics on the pressure dependence of direct band gaps in tetrahedrally bonded semiconductors [45]. In short, for conventional semiconductors with predominantly $sp^3$ hybridization holds that:

(a) States with bonding p-orbital character are almost insensitive to pressure. The best example is the top of the valence band at the Γ point of the Brillouin zone (BZ), which is bonding pure p-type.

(b) On the contrary, antibonding s-states exhibit a strong blue shift with pressure. In most of the cases, the conduction band minimum also at Γ comprises states with antibonding pure s-orbital character. For these reasons, in conventional semiconductors, direct gaps at the BZ center increase with pressure at a rate on the order of 100 meV/GPa [45].

(c) Antibonding p-orbitals are characterized by a much smaller but negative deformation potential as compared with s-states, like conduction band states of the X-valleys at the BZ edge. As a result, the indirect Γ -X gap decreases with pressure at a slower pace of typically ca. -15 meV/GPa [45].

The situation is dramatically different for lead halide perovskites due to the huge spin-orbit splitting of the p-states in heavy atoms like Pb, which leads to a so-called band inversion. Without spin-orbit, the level ordering of s and p atomic orbitals in Pb is similar to that of α-Sn (see Fig. 2.26 in Ref. [59]). Spin-orbit coupling, though, strongly splits the p-states of Pb, pushing the bonding p-states below the energy of the antibonding s-states, such that the atomic orbital character of valence and conduction band states are interchanged, as compared to conventional semiconductors. Relativistic band-structure calculations [60-62] for a pseudo-cubic phase of $MAPbI_3$ predict a direct fundamental gap at the R-point of the BZ with the top of the valence band predominantly composed by antibonding Pb 6s orbitals hybridized with I 5p orbitals and the bottom of the conduction band formed by the antibonding split-off Pb 6p-orbitals. With increasing pressure one thus expects that the top of the valence band shifts up and the bottom of the conduction band shifts slightly down, leading to a reduction of the fundamental gap, as experimentally observed for all lead halide perovskites (see Table 1). As a corollary of the preceding discussion about the gap pressure coefficient, it results that in perovskites the TE term will always lead to a gradual opening of the gap with increasing temperature. Around ambient conditions, where the pressure coefficient as well as thermal expansion coefficient and bulk modulus are temperature independent, the contribution of thermal expansion to the temperature renormalization of the gap is thus a linear function with a positive slope (constant positive derivative). Since the gap temperature coefficient exhibits a clear systematic regarding sign and magnitude across most lead halide perovskites too, we are led to the conclusion that the contribution of the electron-phonon interaction to the gap renormalization is also determined by the bonding/antibonding and atomic orbital character of the electronic states of the band extrema. Such a relationship, although less transparent than for the pressure coefficient, has been elucidated for tetrahedrally bonded semiconductors in the case of a deformation-potential mediated electron-phonon interaction [55, 63, 64]. A similar line of argumentation has been followed to successfully explain the contribution of the EP term to the temperature renormalization of the gap in $MAPbI_3$ [31].

**Table 1.** Linear pressure $dE_g/dP$ and temperature $dE_g/dT$ coefficients of the fundamental band gap, mainly in the case of the tetragonal or cubic phase, stable at ambient conditions, measured for a series of lead halide perovskite materials. The notation SC, PC, TF, and NC stands for single crystal, polycrystal, thin film and nanocrystal, respectively, indicating the structural nature of the samples. For NCs the symbol "<" means average sizes less than 6 nm, ">" means sizes between 6 and 10 nm, and if nothing is indicated then the average NC size is larger than 10 nm. Numbers in parentheses are error bars. The asterisk denotes slope values obtained from only three data points available.

| Material | | $\frac{dE_g}{dP}$ (meV/GPa) | $\frac{dE_g}{dT}$ (meV/K) | Reference |
|---|---|---|---|---|
| $MAPbI_3$ | SC | -50(10) | 0.26(5) | [42] |
| | SC | -65(10)* | | [65] |
| | SC | -50(15)* | | [48] |
| | SC | -70(10)* | | [50] |
| | PC | -62(5) | | [66] |
| | PC | -210(50)* | | [67] |
| | TF | | 0.29(5) | [68] |
| | TF | | 0.24(5) | [69] |
| | TF | | 0.1(1) | [70] |
| | $NC_{>}$ | -43(6) | 0.21(2) | [71] |
| | $NC_{<}$ | -8(7) | 0.19(3) | [71] |
| | NC | | 0.16(6) | [15] |
| $MAPbBr_3$ | SC | -54(5) | 0.24(5) | [9] |
| | SC | | 0.27(5) | [72] |
| | SC | -85(15) | | [73] |
| | PC | -52(8) | | [50] |
| | TF | | 0.14(6) | [69] |
| | TF | | 0.15(8) | [70] |
| | NC | | 0.33(8) | [74] |
| | NC | | 0.36(6) | [18] |
| | NC | | 0.44(10) | [75] |
| $MAPbCl_3$ | PC | -77(5) | | [76] |
| $MA_{0.2}FA_{0.8}PbI_3$ | SC | -54(6) | 0.35(10) | [41] |
| $MA_{0.3}FA_{0.7}PbBr_3$ | TF | -41(5) | 0.20(10) | [39] |
| $MA_{0.3}FA_{0.7}Pb_{0.5}Sn_{0.5}Br_3$ | TF | -82(5) | 0.30(10) | [39] |
| $MA_{0.4}FA_{0.6}PbI_3$ | SC | -55(5) | 0.25(3) | [41] |
| $MA_{0.6}FA_{0.4}PbI_3$ | SC | -58(10) | 0.20(10) | [41] |
| $MA_{0.8}FA_{0.2}PbI_3$ | SC | -46(5) | 0.27(5) | [41] |
| $MA_{0.9}Cs_{0.1}PbI_3$ | SC | -65(15) | 0.11(5) | [32] |
| $MA_{0.95}Cs_{0.05}PbI_3$ | SC | -65(15) | 0.13(5) | [32] |
| $MA_{0.13}EA_{0.87}PbBr_3$ | SC | -61(15) | 2.81(5) | [77] |
| $FAPbI_3$ | SC | | 0.41(5) | [41] |
| | TF | | 0.55(5) | [70] |
| | TF | | 0.38(5) | [78] |
| | NC | | 0.23(6) | [14] |
| $FAPbBr_3$ | SC | | 0.45(5) | [72] |
| | SC | -210(20)* | | [79] |
| | TF | | 0.44(6) | [69] |
| | TF | | 0.40(5) | [70] |
| | NC | | 0.29(6) | [21] |
| $(MHy)PbBr_3$ | SC | -58(15) | -1.97(5) | [77] |
| $CsPbI_3$ | NC | | 0.06(5) | [25] |
| | NC | -20(10) | | [26] |
| | NC | | 0.25(5) | [23] |
| | NC | | 0.25(5) | [80] |
| $CsPbBr_3$ | SC | -72(5) | | [81] |
| | SC | -37(10) | | [82] |
| | SC | | 0.07(10) | [72] |
| | SC | -35(5) | | [83] |
| | SC | -34(15) | 0.31(5) | [77] |
| | $NC_{>}$ | -55(15) | 0.23(5) | This work |
| | NC | | 0.05(5) | [25] |
| | $NC_{>}$ | -40(10) | | [27] |
| | $NC_{<}$ | -15(10) | | [27] |
| | $NC_{>}$ | | 0.18(5) | [84] |
| | $NC_{<}$ | | 0.27(5) | [84] |
| | $NC_{<}$ | | 0.14(6) | [85] |

| | | | |
|---|---|---|---|
| | NC | 0.33(8) | [23] |
| | NC | 0.07(10) | [86] |
| | NC | 0.08(10) | [28] |
| $CsPb(I,Br)_3$ | NC | 0.18(5) | [23] |
| $CsPbCl_3$ | TF | -0.15(10) | [87] |
| | TF | -0.27(10) | [87] |
| | NC | -0.07(5) | [25] |

**CONCLUSIONS**

In this study, we investigated the energy gap behavior in $CsPbBr_3$ NCs as a function of temperature and pressure. We found that the temperature dependence of the band gap in the temperature range from 250 K to 300 K is linear with a positive slope of $2.24(2)\times10^{-4}$ eV/K. By employing high-pressure experiments, the pressure coefficient was acquired. According to these experiments and using the volumetric thermal expansion coefficient and bulk modulus of $CsPbBr_3$ from the literature, contributions of TE and EP to the gap renormalization were disentangled. The obtained value of TE term is $1.3(4)\times10^{-4}$ eV/K and EP contribution is $1.0(3)\times10^{-4}$ eV/K, which demonstrates that both terms have significant and almost equal contributions to the gap renormalization. A single Einstein oscillator model was used to assess the effects of electron-phonon coupling on the temperature renormalization of the gap. This oscillator with positive amplitude, which accounts solely for the coupling with phonons of the inorganic cage, is all what is needed (apart from thermal expansion effects) for an accurate description of the linear gap temperature dependence around room temperature. Interestingly, an exhaustive survey among lead halide perovskites comparing their gap temperature and pressure coefficients shows that this is the most common situation, whereas that of an indirect electron-phonon coupling involving degrees of freedom of the A-site cations is clearly exceptional, i.e. just an anomaly. In this way, we were able to shed light on a fundamental issue, that of the temperature renormalization of the gap of lead halide perovskites, a key information in optoelectronics.

**Supporting Information**

Contains Raman characterization of the $CsPbBr_3$ NCs that shows the common spectra of the cubic phase. Besides, the PL spectra under different pressures are also presented. The maximum of the PL peaks gave the energy gap for each pressure and accordingly the pressure coeficient was obtained.

**Acknowledgements**

The Spanish "Ministerio de Ciencia, Innovación y Universidades" (MICIU) through the Agencia Estatal de Investigación (AEI) is gratefully acknowledged for its support through grant CEX2023-001263-S (MATRANS42) in the framework of the Spanish Severo Ochoa Centre of Excellence program and the AEI/FEDER(UE) grants PID2020-117371RA-I00 (CHIRALPERO), PID2021-128924OB-I00 (ISOSCELLES), PID2022-141956NB-I00 (OUTLIGHT) and TED2021131628A-I00 (MACLEDS). The authors also thank the Catalan agency AGAUR for grant 2021SGR-00444 and the National Network "Red Perovskitas" (MICIU funded). S.F. acknowledges a FPI grant PRE2021-100097 from MICIU and the PhD programme in Materials Science from Universitat Autònoma de Barcelona in which she is enrolled. B.S. acknowledges the support of the Erasmus+ programme of the European Union through the internship project 2022-1-DE01-KA131-HED-000055364 (STREAM 2022). L.P. acknowledges support from the Spanish MICIU through Ramón y Cajal grant (RYC2018-026103-I) and a grant from the Xunta de Galicia (ED431F2021/05). S.G.G. acknowledges support from project CNS2022-135531 (HARDTOP) funded by 14 MCIN/AEI/10.13039/501100011033.

**Data Availability**

All data generated or analyzed during this study are either included in this published article and its supplementary information files or are available from the corresponding author on reasonable request.

**Additional Information**

The authors declare no competing interests.

**References**

1. Dey, A., J. Ye, A. De, E. Debroye, S.K. Ha, E. Bladt, A.S. Kshirsagar, Z. Wang, J. Yin, and Y. Wang, *State of the art and prospects for halide perovskite nanocrystals.* ACS nano, 2021. **15**(7): p. 10775-10981.
2. Dong, H., C. Ran, W. Gao, M. Li, Y. Xia, and W. Huang, *Metal Halide Perovskite for next-generation optoelectronics: progresses and prospects.* ELight, 2023. **3**(1): p. 3.
3. Quan, L.N., B.P. Rand, R.H. Friend, S.G. Mhaisalkar, T.-W. Lee, and E.H. Sargent, *Perovskites for next-generation optical sources.* Chemical reviews, 2019. **119**(12): p. 7444-7477.
4. He, C. and X. Liu, *The rise of halide perovskite semiconductors.* Light: Science & Applications, 2023. **12**(1): p. 15.
5. Hoye, R.L., J. Hidalgo, R.A. Jagt, J.P. Correa-Baena, T. Fix, and J.L. MacManus-Driscoll, *The role of dimensionality on the optoelectronic properties of oxide and halide perovskites, and their halide derivatives.* Advanced Energy Materials, 2022. **12**(4): p. 2100499.
6. Ou, Q., X. Bao, Y. Zhang, H. Shao, G. Xing, X. Li, L. Shao, and Q. Bao, *Band structure engineering in metal halide perovskite nanostructures for optoelectronic applications.* Nano Materials Science, 2019. **1**(4): p. 268-287.
7. Protesescu, L., S. Yakunin, M.I. Bodnarchuk, F. Krieg, R. Caputo, C.H. Hendon, R.X. Yang, A. Walsh, and M.V. Kovalenko, *Nanocrystals of cesium lead halide perovskites (CsPbX3, X= Cl, Br, and I): novel optoelectronic materials showing bright emission with wide color gamut.* Nano letters, 2015. **15**(6): p. 3692-3696.
8. Jiang, H., S. Cui, Y. Chen, and H. Zhong, *Ion exchange for halide perovskite: From nanocrystal to bulk materials.* Nano Select, 2021. **2**(11): p. 2040-2060.
9. Xu, K., L. Pérez-Fidalgo, B.L. Charles, M.T. Weller, M.I. Alonso, and A.R. Goñi, *Using pressure to unravel the structure–dynamic-disorder relationship in metal halide perovskites.* Scientific Reports, 2023. **13**(1): p. 9300.
10. Hoffman, A.E., R.A. Saha, S. Borgmans, P. Puech, T. Braeckevelt, M.B. Roeffaers, J.A. Steele, J. Hofkens, and V. Van Speybroeck, *Understanding the phase transition mechanism in the lead halide perovskite CsPbBr3 via theoretical and experimental GIWAXS and Raman spectroscopy.* Apl Materials, 2023. **11**(4).
11. Frost, J.M. and A. Walsh, *What is moving in hybrid halide perovskite solar cells?* Accounts of chemical research, 2016. **49**(3): p. 528-535.
12. Brennan, M.C., M. Kuno, and S. Rouvimov, *Crystal structure of individual CsPbBr3 perovskite nanocubes.* Inorganic chemistry, 2018. **58**(2): p. 1555-1560.
13. Dirin, D.N., L. Protesescu, D. Trummer, I.V. Kochetygov, S. Yakunin, F. Krumeich, N.P. Stadie, and M.V. Kovalenko, *Harnessing defect-tolerance at the nanoscale: highly luminescent lead halide perovskite nanocrystals in mesoporous silica matrixes.* Nano letters, 2016. **16**(9): p. 5866-5874.
14. Diroll, B.T., P. Guo, and R.D. Schaller, *Unique optical properties of methylammonium lead iodide nanocrystals below the bulk tetragonal-orthorhombic phase transition.* Nano letters, 2018. **18**(2): p. 846-852.
15. Shi, Z.-F., Y. Li, S. Li, H.-F. Ji, L.-Z. Lei, D. Wu, T.-T. Xu, J.-M. Xu, Y.-T. Tian, and X.-J. Li, *Polarized emission effect realized in CH 3 NH 3 PbI 3 perovskite nanocrystals.* Journal of Materials Chemistry C, 2017. **5**(34): p. 8699-8706.
16. Ijaz, P., M. Imran, M.M. Soares, H.C. Tolentino, B. Martín-García, C. Giannini, I. Moreels, L. Manna, and R. Krahne, *Composition-, size-, and surface functionalization-dependent optical properties of lead bromide perovskite nanocrystals.* The Journal of Physical Chemistry Letters, 2020. **11**(6): p. 2079-2085.
17. Rubino, A., M. Anaya, J.F. Galisteo-López, T.C. Rojas, M.E. Calvo, and H. Míguez, *Highly efficient and environmentally stable flexible color converters based on confined CH3NH3PbBr3 nanocrystals.* ACS applied materials & interfaces, 2018. **10**(44): p. 38334-38340.

18. Woo, H.C., J.W. Choi, J. Shin, S.-H. Chin, M.H. Ann, and C.-L. Lee, *Temperature-dependent photoluminescence of CH3NH3PbBr3 perovskite quantum dots and bulk counterparts.* The journal of physical chemistry letters, 2018. **9**(14): p. 4066-4074.
19. Fang, H.H., L. Protesescu, D.M. Balazs, S. Adjokatse, M.V. Kovalenko, and M.A. Loi, *Exciton recombination in formamidinium lead triiodide: nanocrystals versus thin films.* small, 2017. **13**(32): p. 1700673.
20. Fu, M., P. Tamarat, J.-B. Trebbia, M.I. Bodnarchuk, M.V. Kovalenko, J. Even, and B. Lounis, *Unraveling exciton–phonon coupling in individual FAPbI3 nanocrystals emitting near-infrared single photons.* Nature communications, 2018. **9**(1): p. 3318.
21. Ghosh, S., Q. Shi, B. Pradhan, P. Kumar, Z. Wang, S. Acharya, S.K. Pal, T. Pullerits, and K.J. Karki, *Phonon coupling with excitons and free carriers in formamidinium lead bromide perovskite nanocrystals.* The journal of physical chemistry letters, 2018. **9**(15): p. 4245-4250.
22. Pfingsten, O., J. Klein, L. Protesescu, M.I. Bodnarchuk, M.V. Kovalenko, and G. Bacher, *Phonon interaction and phase transition in single formamidinium lead bromide quantum dots.* Nano letters, 2018. **18**(7): p. 4440-4446.
23. Lee, S.M., C.J. Moon, H. Lim, Y. Lee, M.Y. Choi, and J. Bang, *Temperature-dependent photoluminescence of cesium lead halide perovskite quantum dots: splitting of the photoluminescence peaks of $CsPbBr_3$ and $CsPb(Br/I)_3$ quantum dots at low temperature.* The Journal of Physical Chemistry C, 2017. **121**(46): p. 26054-26062.
24. Liu, A., L.G. Bonato, F. Sessa, D.B. Almeida, E. Isele, G. Nagamine, L.F. Zagonel, A.F. Nogueira, L.A. Padilha, and S.T. Cundiff, *Effect of dimensionality on the optical absorption properties of CsPbI3 perovskite nanocrystals.* The Journal of chemical physics, 2019. **151**(19).
25. Saran, R., A. Heuer-Jungemann, A.G. Kanaras, and R.J. Curry, *Giant bandgap renormalization and exciton–phonon scattering in perovskite nanocrystals.* Advanced Optical Materials, 2017. **5**(17): p. 1700231.
26. Vukovic, O., G. Folpini, E.L. Wong, L. Leoncino, G. Terraneo, M.D. Albaqami, A. Petrozza, and D. Cortecchia, *Structural effects on the luminescence properties of CsPbI 3 nanocrystals.* Nanoscale, 2023. **15**(12): p. 5712-5719.
27. Beimborn, J.C., L.R. Walther, K.D. Wilson, and J.M. Weber, *Size-dependent pressure-response of the photoluminescence of CsPbBr3 nanocrystals.* The Journal of Physical Chemistry Letters, 2020. **11**(5): p. 1975-1980.
28. Strandell, D.P. and P. Kambhampati, *The temperature dependence of the photoluminescence of CsPbBr3 nanocrystals reveals phase transitions and homogeneous linewidths.* The Journal of Physical Chemistry C, 2021. **125**(49): p. 27504-27508.
29. Yin, Y., Y. Liu, G. Cao, Z. Lv, H. Zong, Y. Cheng, Q. Dong, C. Liu, M. Li, and B. Zhang, *Optical properties and mechanical induced phase transition of CsPb2Br5 and CsPbBr3 nanocrystals.* Journal of Alloys and Compounds, 2023. **947**: p. 169439.
30. Lohar, A.A., A. Shinde, R. Gahlaut, A. Sagdeo, and S. Mahamuni, *Enhanced photoluminescence and stimulated emission in CsPbCl3 nanocrystals at low temperature.* The Journal of Physical Chemistry C, 2018. **122**(43): p. 25014-25020.
31. Francisco-López, A., B. Charles, O.J. Weber, M.I. Alonso, M. Garriga, M. Campoy-Quiles, M.T. Weller, and A.R. Goñi, *Equal footing of thermal expansion and electron–phonon interaction in the temperature dependence of lead halide perovskite band gaps.* The Journal of Physical Chemistry Letters, 2019. **10**(11): p. 2971-2977.
32. Pérez-Fidalgo, L., K. Xu, B.L. Charles, P.F. Henry, M.T. Weller, M.I. Alonso, and A.R. Goñi, *Anomalous Electron–Phonon Coupling in Cesium-Substituted Methylammonium Lead Iodide Perovskites.* The Journal of Physical Chemistry C, 2023. **127**(46): p. 22817-22826.

33. Ghosh, S., D. Rana, B. Pradhan, P. Donfack, J. Hofkens, and A. Materny, *Vibrational study of lead bromide perovskite materials with variable cations based on Raman spectroscopy and density functional theory.* Journal of Raman Spectroscopy, 2021. **52**(12): p. 2338-2347.
34. Leguy, A.M., A.R. Goñi, J.M. Frost, J. Skelton, F. Brivio, X. Rodríguez-Martínez, O.J. Weber, A. Pallipurath, M.I. Alonso, and M. Campoy-Quiles, *Dynamic disorder, phonon lifetimes, and the assignment of modes to the vibrational spectra of methylammonium lead halide perovskites.* Physical Chemistry Chemical Physics, 2016. **18**(39): p. 27051-27066.
35. Angel, R.J., M. Bujak, J. Zhao, G.D. Gatta, and S.D. Jacobsen, *Effective hydrostatic limits of pressure media for high-pressure crystallographic studies.* Journal of Applied Crystallography, 2007. **40**(1): p. 26-32.
36. Mao, H., J.-A. Xu, and P. Bell, *Calibration of the ruby pressure gauge to 800 kbar under quasi-hydrostatic conditions.* Journal of Geophysical Research: Solid Earth, 1986. **91**(B5): p. 4673-4676.
37. López, C.A., C. Abia, M.C. Alvarez-Galván, B.-K. Hong, M.V. Martínez-Huerta, F. Serrano-Sánchez, F. Carrascoso, A.s. Castellanos-Gómez, M.T. Fernández-Díaz, and J.A. Alonso, *Crystal structure features of CsPbBr3 perovskite prepared by mechanochemical synthesis.* ACS omega, 2020. **5**(11): p. 5931-5938.
38. Boehme, S.C., M.I. Bodnarchuk, M. Burian, F. Bertolotti, I. Cherniukh, C. Bernasconi, C. Zhu, R. Erni, H. Amenitsch, and D. Naumenko, *Strongly confined CsPbBr3 quantum dots as quantum emitters and building blocks for rhombic superlattices.* ACS nano, 2023. **17**(3): p. 2089-2100.
39. Zhang, H., Z. Bi, Z. Zhai, H. Gao, Y. Liu, M. Jin, M. Ye, X. Li, H. Liu, and Y. Zhang, *Revealing Unusual Bandgap Shifts with Temperature and Bandgap Renormalization Effect in Phase-Stabilized Metal Halide Perovskite Thin Films.* Advanced Functional Materials, 2024. **34**(9): p. 2302214.
40. Steele, J.A., M. Lai, Y. Zhang, Z. Lin, J. Hofkens, M.B. Roeffaers, and P. Yang, *Phase transitions and anion exchange in all-inorganic halide perovskites.* Accounts of Materials Research, 2020. **1**(1): p. 3-15.
41. Francisco-López, A., B. Charles, M.I. Alonso, M. Garriga, M. Campoy-Quiles, M.T. Weller, and A.R. Goñi, *Phase diagram of methylammonium/formamidinium lead iodide perovskite solid solutions from temperature-dependent photoluminescence and Raman spectroscopies.* The Journal of Physical Chemistry C, 2020. **124**(6): p. 3448-3458.
42. Francisco-López, A.n., B. Charles, O.J. Weber, M.I. Alonso, M. Garriga, M. Campoy-Quiles, M.T. Weller, and A.R. Goñi, *Pressure-induced locking of methylammonium cations versus amorphization in hybrid lead iodide perovskites.* The Journal of Physical Chemistry C, 2018. **122**(38): p. 22073-22082.
43. Francisco-López, A., B. Charles, M.I. Alonso, M. Garriga, M.T. Weller, and A.R. Goñi, *Photoluminescence of Bound-Exciton Complexes and Assignment to Shallow Defects in Methylammonium/Formamidinium Lead Iodide Mixed Crystals.* Advanced Optical Materials, 2021. **9**(18): p. 2001969.
44. Wojdyr, M., *Fityk: a general-purpose peak fitting program.* Journal of applied crystallography, 2010. **43**(5): p. 1126-1128.
45. Goni, A. and K. Syassen, *Optical properties of semiconductors under pressure*, in *Semiconductors and Semimetals*. 1998, Elsevier. p. 247-425.
46. Ghosh, D., A. Aziz, J.A. Dawson, A.B. Walker, and M.S. Islam, *Putting the squeeze on lead iodide perovskites: pressure-induced effects to tune their structural and optoelectronic behavior.* Chemistry of Materials, 2019. **31**(11): p. 4063-4071.
47. Ghosh, D., P. Walsh Atkins, M.S. Islam, A.B. Walker, and C. Eames, *Good vibrations: locking of octahedral tilting in mixed-cation iodide perovskites for solar cells.* ACS Energy Letters, 2017. **2**(10): p. 2424-2429.

48. Jaffe, A., Y. Lin, C.M. Beavers, J. Voss, W.L. Mao, and H.I. Karunadasa, *High-pressure single-crystal structures of 3D lead-halide hybrid perovskites and pressure effects on their electronic and optical properties.* ACS central science, 2016. **2**(4): p. 201-209.
49. Jaffe, A., Y. Lin, and H.I. Karunadasa, *Halide perovskites under pressure: accessing new properties through lattice compression.* ACS Energy Letters, 2017. **2**(7): p. 1549-1555.
50. Kong, L., G. Liu, J. Gong, Q. Hu, R.D. Schaller, P. Dera, D. Zhang, Z. Liu, W. Yang, and K. Zhu, *Simultaneous band-gap narrowing and carrier-lifetime prolongation of organic–inorganic trihalide perovskites.* Proceedings of the National Academy of Sciences, 2016. **113**(32): p. 8910-8915.
51. Liu, G., L. Kong, W. Yang, and H.-k. Mao, *Pressure engineering of photovoltaic perovskites.* Materials Today, 2019. **27**: p. 91-106.
52. Celeste, A. and F. Capitani, *Hybrid perovskites under pressure: Present and future directions.* Journal of Applied Physics, 2022. **132**(22).
53. Haeger, T., R. Heiderhoff, and T. Riedl, *Thermal properties of metal-halide perovskites.* Journal of Materials Chemistry C, 2020. **8**(41): p. 14289-14311.
54. Ezzeldien, M., S. Al-Qaisi, Z. Alrowaili, M. Alzaid, E. Maskar, A. Es-Smairi, T.V. Vu, and D. Rai, *Electronic and optical properties of bulk and surface of CsPbBr3 inorganic halide perovskite a first principles DFT 1/2 approach.* Scientific Reports, 2021. **11**(1): p. 20622.
55. Gopalan, S., P. Lautenschlager, and M. Cardona, *Temperature dependence of the shifts and broadenings of the critical points in GaAs.* Physical Review B, 1987. **35**(11): p. 5577.
56. Bhosale, J., A. Ramdas, A. Burger, A. Muñoz, A. Romero, M. Cardona, R. Lauck, and R. Kremer, *Temperature dependence of band gaps in semiconductors: Electron-phonon interaction.* Physical Review B, 2012. **86**(19): p. 195208.
57. Göbel, A., T. Ruf, M. Cardona, C. Lin, J. Wrzesinski, M. Steube, K. Reimann, J.-C. Merle, and M. Joucla, *Effects of the isotopic composition on the fundamental gap of CuCl.* Physical Review B, 1998. **57**(24): p. 15183.
58. Serrano, J., C. Schweitzer, C. Lin, K. Reimann, M. Cardona, and D. Fröhlich, *Electron-phonon renormalization of the absorption edge of the cuprous halides.* Physical Review B, 2002. **65**(12): p. 125110.
59. Yu, P.Y., M. Cardona, P.Y. Yu, and M. Cardona, *Electronic band structures.* Fundamentals of Semiconductors: Physics and Materials Properties, 2010: p. 17-106.
60. Even, J., L. Pedesseau, C. Katan, M. Kepenekian, J.-S. Lauret, D. Sapori, and E. Deleporte, *Solid-state physics perspective on hybrid perovskite semiconductors.* The Journal of Physical Chemistry C, 2015. **119**(19): p. 10161-10177.
61. Frost, J.M., K.T. Butler, F. Brivio, C.H. Hendon, M. Van Schilfgaarde, and A. Walsh, *Atomistic origins of high-performance in hybrid halide perovskite solar cells.* Nano letters, 2014. **14**(5): p. 2584-2590.
62. Giorgi, G., J.-I. Fujisawa, H. Segawa, and K. Yamashita, *Small photocarrier effective masses featuring ambipolar transport in methylammonium lead iodide perovskite: a density functional analysis.* The journal of physical chemistry letters, 2013. **4**(24): p. 4213-4216.
63. Allen, P. and M. Cardona, *Theory of the temperature dependence of the direct gap of germanium.* Physical Review B, 1981. **23**(4): p. 1495.
64. Lautenschlager, P., P.B. Allen, and M. Cardona, *Temperature dependence of band gaps in Si and Ge.* Phys Rev B Condens Matter, 1985. **31**(4): p. 2163-2171.
65. Szafranski, M. and A. Katrusiak, *Mechanism of pressure-induced phase transitions, amorphization, and absorption-edge shift in photovoltaic methylammonium lead iodide.* The journal of physical chemistry letters, 2016. **7**(17): p. 3458-3466.

66. Wang, T., B. Daiber, J.M. Frost, S.A. Mann, E.C. Garnett, A. Walsh, and B. Ehrler, *Indirect to direct bandgap transition in methylammonium lead halide perovskite.* Energy & Environmental Science, 2017. **10**(2): p. 509-515.
67. Jiang, S., Y. Fang, R. Li, H. Xiao, J. Crowley, C. Wang, T.J. White, W.A. Goddard III, Z. Wang, and T. Baikie, *Pressure-dependent polymorphism and band-gap tuning of methylammonium lead iodide perovskite.* Angewandte Chemie International Edition, 2016. **55**(22): p. 6540-6544.
68. Milot, R.L., G.E. Eperon, H.J. Snaith, M.B. Johnston, and L.M. Herz, *Temperature-dependent charge-carrier dynamics in CH3NH3PbI3 perovskite thin films.* Advanced Functional Materials, 2015. **25**(39): p. 6218-6227.
69. Dar, M.I., G. Jacopin, S. Meloni, A. Mattoni, N. Arora, A. Boziki, S.M. Zakeeruddin, U. Rothlisberger, and M. Grätzel, *Origin of unusual bandgap shift and dual emission in organic-inorganic lead halide perovskites.* Science advances, 2016. **2**(10): p. e1601156.
70. Wright, A.D., C. Verdi, R.L. Milot, G.E. Eperon, M.A. Pérez-Osorio, H.J. Snaith, F. Giustino, M.B. Johnston, and L.M. Herz, *Electron–phonon coupling in hybrid lead halide perovskites.* Nature communications, 2016. **7**(1): p. 11755.
71. Rubino, A., A. Francisco-López, A.J. Barker, A. Petrozza, M.E. Calvo, A.R. Goñi, and H. Míguez, *Disentangling Electron–Phonon Coupling and Thermal Expansion Effects in the Band Gap Renormalization of Perovskite Nanocrystals.* The Journal of Physical Chemistry Letters, 2020. **12**(1): p. 569-575.
72. Mannino, G., I. Deretzis, E. Smecca, A. La Magna, A. Alberti, D. Ceratti, and D. Cahen, *Temperature-dependent optical band gap in CsPbBr3, MAPbBr3, and FAPbBr3 single crystals.* The journal of physical chemistry letters, 2020. **11**(7): p. 2490-2496.
73. Liang, A., R. Turnbull, C. Popescu, I. Fernandez-Guillen, R. Abargues, P.P. Boix, and D. Errandonea, *Pressure-induced phase transition versus amorphization in hybrid methylammonium lead bromide perovskite.* The Journal of Physical Chemistry C, 2023. **127**(26): p. 12821-12826.
74. Li, J., Z. Guo, S. Xiao, Y. Tu, T. He, and W. Zhang, *Optimizing optical properties of hybrid core/shell perovskite nanocrystals.* Inorganic Chemistry Frontiers, 2022. **9**(12): p. 2980-2986.
75. Sadhukhan, P., A. Pradhan, S. Mukherjee, P. Sengupta, A. Roy, S. Bhunia, and S. Das, *Low temperature excitonic spectroscopy study of mechano-synthesized hybrid perovskite.* Applied Physics Letters, 2019. **114**(13).
76. Wang, L., K. Wang, G. Xiao, Q. Zeng, and B. Zou, *Pressure-induced structural evolution and band gap shifts of organometal halide perovskite-based methylammonium lead chloride.* The journal of physical chemistry letters, 2016. **7**(24): p. 5273-5279.
77. Huang, X., X. Li, Y. Tao, S. Guo, J. Gu, H. Hong, Y. Yao, Y. Guan, Y. Gao, and C. Li, *Understanding electron–phonon interactions in 3D lead halide perovskites from the stereochemical expression of 6s2 lone pairs.* Journal of the American Chemical Society, 2022. **144**(27): p. 12247-12260.
78. Wright, A.D., G. Volonakis, J. Borchert, C.L. Davies, F. Giustino, M.B. Johnston, and L.M. Herz, *Intrinsic quantum confinement in formamidinium lead triiodide perovskite.* Nature Materials, 2020. **19**(11): p. 1201-1206.
79. Wang, L., K. Wang, and B. Zou, *Pressure-induced structural and optical properties of organometal halide perovskite-based formamidinium lead bromide.* The journal of physical chemistry letters, 2016. **7**(13): p. 2556-2562.
80. Gau, D.L., I. Galain, I. Aguiar, and R.E. Marotti, *Origin of photoluminescence and experimental determination of exciton binding energy, exciton-phonon interaction, and urbach energy in γ-CsPbI3 nanoparticles.* Journal of Luminescence, 2023. **257**: p. 119765.

81. Gong, J., H. Zhong, C. Gao, J. Peng, X. Liu, Q. Lin, G. Fang, S. Yuan, Z. Zhang, and X. Xiao, *Pressure-Induced Indirect-Direct Bandgap Transition of $CsPbBr_3$ Single Crystal and Its Effect on Photoluminescence Quantum Yield.* Advanced Science, 2022. **9**(29): p. 2201554.
82. Bulyk, L.-I., T. Demkiv, O. Antonyak, Y.M. Chornodolskyy, R. Gamernyk, A. Suchocki, and A. Voloshinovskii, *Pressure influence on excitonic luminescence of $CsPbBr_3$ perovskite.* Dalton Transactions, 2023. **52**(45): p. 16712-16719.
83. Zhang, L., Q. Zeng, and K. Wang, *Pressure-induced structural and optical properties of inorganic halide perovskite $CsPbBr_3$.* The journal of physical chemistry letters, 2017. **8**(16): p. 3752-3758.
84. Li, J., X. Yuan, P. Jing, J. Li, M. Wei, J. Hua, J. Zhao, and L. Tian, *Temperature-dependent photoluminescence of inorganic perovskite nanocrystal films.* RSC advances, 2016. **6**(82): p. 78311-78316.
85. Shinde, A., R. Gahlaut, and S. Mahamuni, *Low-temperature photoluminescence studies of $CsPbBr_3$ quantum dots.* The Journal of Physical Chemistry C, 2017. **121**(27): p. 14872-14878.
86. Zhang, X., X. Gao, G. Pang, T. He, G. Xing, and R. Chen, *Effects of material dimensionality on the optical properties of $CsPbBr_3$ nanomaterials.* The Journal of Physical Chemistry C, 2019. **123**(47): p. 28893-28897.
87. Xu, F., H. Wei, Y. Wu, Y. Zhou, J. Li, and B. Cao, *Nonmonotonic temperature-dependent bandgap change of $CsPbCl_3$ films induced by optical phonon scattering.* Journal of Luminescence, 2023. **257**: p. 119736.

Supporting information

# Absence of Anomalous Electron-Phonon Coupling in the Temperature Renormalization of the Gap of $CsPbBr_3$ Nanocrystals

Shima Fasahat[1], Benedikt Schäfer[1], Kai Xu[1], Nadesh Fiuza-Maneiro[2], Sergio Gómez-Graña[2], M. Isabel Alonso[1], Lakshminarayana Polavarapu[2], Alejandro R. Goñi[1,3*]

[1]Institut de Ciencia de Materials de Barcelona, ICMAB-CSIC, Campus UAB, 08193 Bellaterra, Spain
[2]CINBIO, Universidade de Vigo, Materials Chemistry and Physics Group, Dept. of Physical Chemistry, Campus Universitario Lagoas Marcosende, 36310 Vigo, Spain
[3]ICREA, Passeig Lluís Companys 23, 08010 Barcelona, Spain

A representative Raman spectrum of $CsPbBr_3$ NCs recorded at ambient pressure and temperature conditions is shown in Fig. S1. Typical of the cubic phase, for which the Cs dynamics is known to be fully unfolded, is the steep increase in Raman intensity for very small Raman shifts, due to the presence of a "zero-shift" peak. This peak arises from incoherent scattering due to fluctuations in the Raman susceptibility as a consequence of the unleashed Cs dynamics and its effect on the inorganic cage phonons [1]. Besides, three broad bands can be barely resolved on top of the steep slope, in the spectral range of the inorganic cage phonons below 200 $cm^{-1}$. The strong broadening of the Raman peaks is additional evidence of the presence of dynamic disorder, a characteristic feature of the cubic phase [2,3].

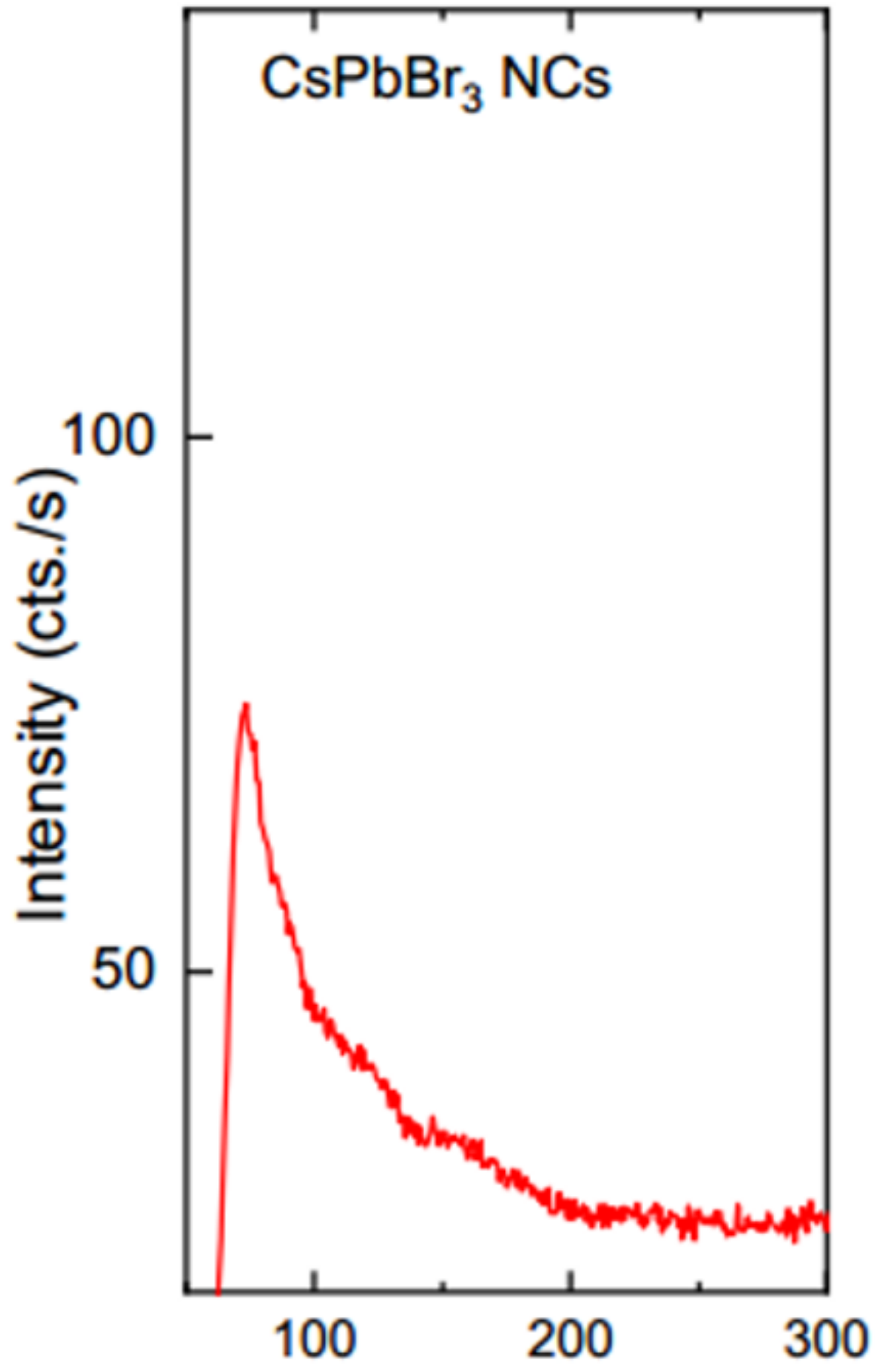

Figure S1: Room-temperature Raman spectrum of $CsPbBr_3$ NCs.

Representative photoluminescence (PL) spectra of $CsPbBr_3$ NCs measured at different pressures at room temperature are shown in Fig. S2. Initially, the main PL peak shifts to the red with increasing pressure. The line shape changes observed in the PL spectra starting at 0.91 GPa are a consequence of the occurrence of a pressure-induced first-order phase transition [4]. Within a certain range around the phase transition pressure, the coexistence of the two crystal phases is observed (double PL peak). To calculate the TE term of the gap temperature dependence, the pressure coefficient of the gap is obtained only from the PL peak positions of the spectra from the phase stable at ambient conditions.

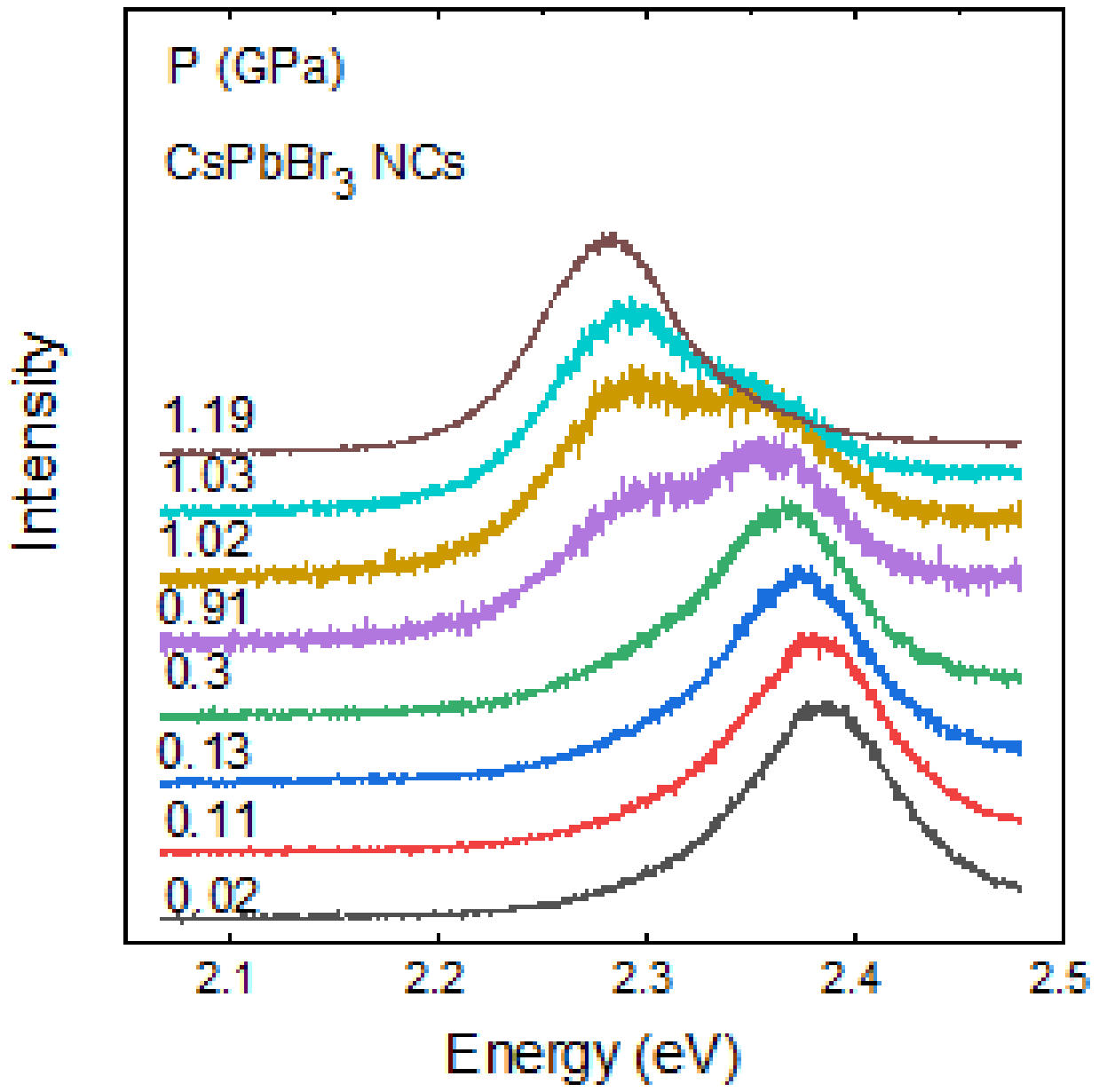

Figure S2: PL spectra of CsPbBr3 NCs under different pressures.

**References**

[1] Yaffe, O.; Guo, Y.; Tan, L. Z.; Egger, D. A.; Hull, T.; Stoumpos, C. C.; Zheng, F.; Heinz, T. F.; Kronik, L.; Kanatzidis, M. G.; et al. Local Polar Fluctuations in Lead Halide Perovskite Crystals. *Phys. Rev. Lett.* **2017**, 118, 136001.

[2] Leguy, A. M. A.; Goñi, A. R.; Frost, J. M.; Skelton, J.; Brivio, F.; Rodríguez-Martínez, X.; Weber, O. J.; Pallipurath, A.; Alonso, M. I.; Campoy-Quiles, M.; Weller, M. T.; Nelson, J.; Walsh, A.; Barnes, P. R. F. Dynamic Disorder, Phonon Lifetimes, and the Assignment of Modes to the Vibrational Spectra of Methylammonium Lead Halide Perovskites. *Phys. Chem. Chem. Phys.* **2016**, 18, 27051-27066.

[3] Goñi, A. R. Raman Linewidths as a Probe of Lattice Anharmonicity and Dynamic Disorder in Metal Halide Perovskites. *Asian J. Phys.* **2024**, 33, 29-38.

[4] Beimborn, J.C.; Walther, L.R.; Wilson, K.D.; Weber, J.M. Size-dependent pressure-response of the photoluminescence of $CsPbBr_3$ nanocrystals. *J. Phys. Chem. Lett.* **2020,** 11, 1975-1980.